\documentclass[conference]{IEEEtran}
\usepackage{amsmath,amssymb,amsfonts}
\usepackage{algorithmic}
\usepackage{algorithm} 
\usepackage{graphicx}
\usepackage{textcomp}
\usepackage{xcolor}
\usepackage{stfloats}
\usepackage{hyperref}
\usepackage{tikz}
\usetikzlibrary{shapes,arrows,positioning,fit,backgrounds,decorations.pathreplacing,calligraphy}
\usepackage{pgfplots}
\pgfplotsset{compat=1.18}
\usepackage{comment}

\begin{document}

\title{Composable Attestation: A Generalized Framework for Continuous and Incremental Trust in AI-Driven Distributed Systems}

\author{
\IEEEauthorblockN{Sheng Sun}
\IEEEauthorblockA{\textit{Dell Canada}\\
Ottawa, Canada\\
robert.sun@dell.com}
\and
\IEEEauthorblockN{Sarah Evans}
\IEEEauthorblockA{\textit{Dell Technologies Inc}\\
Denver, USA \\
sarah\_evans1@dell.com}
}

\maketitle

\begin{abstract}
This paper presents composable attestation as a generalized cryptographic framework for Continuous and Incremental Trust in  Distributed Systems,such as Artificial Intelligence (AI) computation, and Open Source Software (OSS) supply chain verification. We establish a rigorous mathematical foundation which is defining core properties of such attestation systems: composability, order independence, transitivity, determinism, inclusion, and dynamic component verification. In contrast to traditional attestation methodologies relying on monolithic verification \cite{pfi_attestation_overview}, composable attestation facilitates modular,   scalable, and cryptographically secured integrity verification adaptable to evolving system configurations. This work introduces generalized attestation proof generation and verification functions, implementable via a variety of cryptographic constructions, in which Merkle trees \cite{merkle1987} plays vital role in constructing the composable attestation proof.  Alternative constructions, including accumulator-based schemes \cite{benaloh1990} and multi-signature approaches \cite{boneh2001short}, are also explored, each presenting distinct trade-offs in performance, security, and functionality. Formal analysis demonstrates the adherence of these implementations to the fundamental properties . The framework's utility extends to applications such as secure AI model integrity verification \cite{zhang2021attestation}, federated learning \cite{chen2019shielding}, and runtime trust assurance. The concept of attestation inclusion is introduced, permitting incremental integration of new components without necessitating full system re-attestation. This generalized approach reinforce trust in AI computation and broader distributed computing environments through cryptographically verifiable proof mechanisms, building upon foundational concepts of bootstrapping trust \cite{parno2011bootstrapping}.
\end{abstract}

\section{Introduction}
Attestation generates cryptographic evidence of a system's state, configuration, or behavior, thereby establishing trust in distributed computing environments \cite{pfi_attestation_overview}. Composable attestation refines this concept by enabling structured, scalable, and modular proofs of integrity and trustworthiness. Unlike traditional attestation methods, which often employs monolithic or point-of-time verification, composable attestation permits individual system components to be attested independently while maintaining a globally verifiable proof of the entire system's integrity. The principles of composability in security protocols are crucial for building complex trustworthy systems \cite{cai2021universal}.

Modern distributed systems, particularly those engaged in AI computation, comprise numerous components that may evolve autonomously and dynamically. These elements encompass hardware (e.g., CPUs, GPUs, TPMs \cite{intel_sgx_attestation, arm_cca_attestation}), software dependencies, model parameters, and runtime configurations, all susceptible to frequent modifications during standard operation. Conventional attestation techniques often encounter challenges with this dynamics, typically mandating complete system re-attestation upon any component alteration. Early systems for code integrity already highlighted the importance of such verification \cite{kochervc}.

This paper introduces a generalized formal model for composable attestation, emphasizing several essential properties. \textbf{Composability} refers to the capacity to amalgamate attestations of subcomponents into a verifiable attestation of the entire system. \textbf{Order Independence} ensures that the attestation result remains consistent regardless of the sequence in which components are processed. \textbf{Transitivity} signifies the ability to preserve attestation validity when augmenting the system with new components. \textbf{Determinism} guarantees that repeated attestations of an identical component set yield identical proofs. \textbf{Inclusion} allows for the incremental addition of a new component into the attestation structure without demanding full re-attestation, a concept seen in dynamic data structures \cite{naor1999merkle, camenisch2002dynamic}. Finally, \textbf{Dynamic Component Verification} captures attestation across execution units and verifies updates without disrupting ongoing computations. These properties aim to support more abstract, property-based attestation goals \cite{sadeghi2015property}.

The remainder of this paper is structured as follows. Section II delineates the mathematical foundations of composable attestation, incorporating formal definitions of attestation proofs and verification functions. Section III investigates various cryptographic constructions for implementing composable attestation. Section IV analyzes how these constructions fulfill the key mathematical properties. Section V explores applications in AI computation and distributed systems, building on surveys of existing techniques \cite{eldib2020comprehensive}. Section VI concludes the paper and proposes avenues for future research.

\section{Mathematical Foundations of Composable Attestation}

\subsection{Formal Definition of Attestation Proofs}
Let $S = \{C_1, C_2, \ldots, C_n\}$ denote a set of components within a system, where each $C_i$ represents a distinct computational or data entity subject to attestation.

The \textbf{attestation proof generation function}, $A$, is defined as:
\begin{equation}
A: S \rightarrow \pi
\end{equation}
where $\pi$ is the attestation proof derived from the set $S$, uniquely representing the integrity of its constituent components.

The \textbf{verification function}, $V$, is defined as:
\begin{equation}
V: (\pi, S) \rightarrow \{0, 1\}
\end{equation}
where $V(\pi, S) = 1$ if and only if $\pi$ constitutes a valid attestation proof for $S$; otherwise, it returns $0$. Secure communication channels, often established via mechanisms like SKEME \cite{krawczyk2000skeme}, are assumed for the exchange of proofs and challenges.

\subsection{Properties of Composable Attestation}
For effective composable attestation, satisfaction of the following key properties is imperative:

\textbf{Composability}: If $S_1$ and $S_2$ are subsets of $S$, then the attestation of their union can be related to their individual attestations:
\begin{equation}
A(S_1 \cup S_2) = A(S_1) \oplus A(S_2)
\end{equation}
where $\oplus$ represents a cryptographic aggregation operation, the specifics of which depend on the underlying cryptographic construction. Rigorous frameworks like Universal Composability help in formally defining and analyzing such properties \cite{cai2021universal}.

\textbf{Order Independence}: The attestation proof must be invariant to the order in which components are processed:
\begin{equation}
A(\{C_1, C_2, \ldots, C_n\}) = A(\{C_{\sigma(1)}, C_{\sigma(2)}, \ldots, C_{\sigma(n)}\})
\end{equation}
for any permutation $\sigma$ of the indices $\{1, 2, \ldots, n\}$. This property ensures consistency regardless of the component processing sequence.

\textbf{Transitivity}: Given an extended system $S' = S \cup S_3$, verification consistency is maintained:
\begin{equation}
V(A(S'), S') = 1 \implies V(A(S), S) = 1
\end{equation}
This ensures that the attestation proof of a new system configuration preserves the validity of its prior components' attestations.

\textbf{Determinism}: If the system state remains unaltered, the attestation proof must be identical upon repeated evaluations:
\begin{equation}
A(S)_{\text{eval1}} = A(S)_{\text{eval2}}
\end{equation}
This property ensures stability and consistency in proof generation.

\textbf{Inclusion}: If a new component $C'$ is introduced to an already attested system $S$, it should be feasible to generate an updated proof $A(S')$ incrementally, without recomputing all prior attestations:
\begin{equation}
A(S') = A(S \cup \{C'\}) = A(S) \oplus A(\{C'\})
\end{equation}
This significantly enhances scalability by obviating costly recomputation of the entire attestation structure, drawing on principles from dynamic cryptographic structures \cite{camenisch2002dynamic}.

\textbf{Dynamic Attestation}: If components undergo dynamic updates during execution, incremental attestation ensures that verification remains valid without necessitating full system resets or re-attestations.

\section{Cryptographic Constructions for Composable Attestation}

While multiple cryptographic constructions can satisfy the properties required for composable attestation, this section will focus on   Merkle tree-based constructions \cite{merkle1987}. We also explore cryptographic accumulators \cite{benaloh1990} and multisignature schemes \cite{boldyreva2003threshold}. Each offers distinct advantages and trade-offs. The security of these often relies on foundational cryptographic assumptions, such as those analyzed in the random oracle model for hash functions \cite{bellare1993random}.

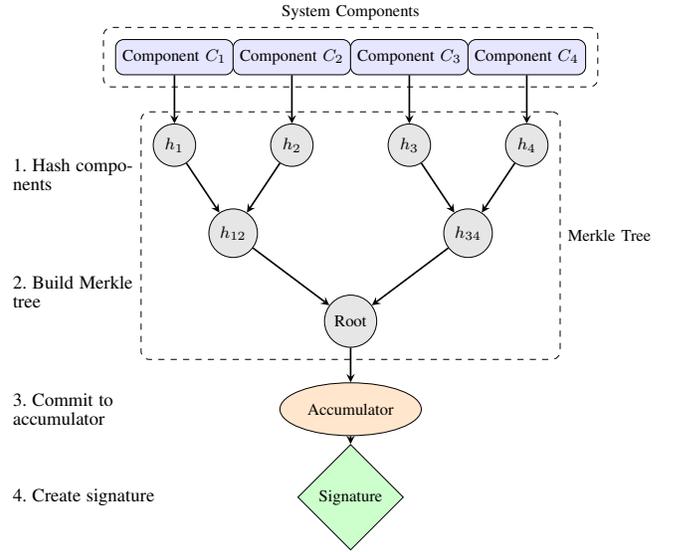
\begin{figure}[t]
\centering
\resizebox{\columnwidth}{!}{
\begin{tikzpicture}[
    node distance=0.5cm,
    box/.style={draw, rectangle, minimum width=1.5cm, minimum height=0.7cm, align=center, font=\tiny},
    component/.style={draw, rectangle, rounded corners, minimum width=1.3cm, minimum height=0.6cm, align=center, fill=blue!10, font=\footnotesize},
    hash/.style={draw, circle, minimum size=0.6cm, fill=gray!20, font=\footnotesize},
    accumulator/.style={draw, ellipse, minimum width=1.8cm, minimum height=0.9cm, fill=orange!20, font=\footnotesize},
    signature/.style={draw, diamond, minimum width=1.3cm, minimum height=0.9cm, fill=green!20, font=\footnotesize},
    arrow/.style={->, >=stealth, thick}
]

\node[component] (c1) at (0,0) {Component $C_1$};
\node[component] (c2) at (2,0) {Component $C_2$};
\node[component] (c3) at (4,0) {Component $C_3$};
\node[component] (c4) at (6,0) {Component $C_4$};

\node[hash] (h1) at (0,-1.5) {$h_1$};
\node[hash] (h2) at (2,-1.5) {$h_2$};
\node[hash] (h3) at (4,-1.5) {$h_3$};
\node[hash] (h4) at (6,-1.5) {$h_4$};

\node[hash] (h12) at (1,-3) {$h_{12}$};
\node[hash] (h34) at (5,-3) {$h_{34}$};

\node[hash] (root) at (3,-4.5) {Root};

\node[accumulator] (acc) at (3,-6) { Accumulator};

\node[signature] (sig) at (3,-7.5) {Signature};

\draw[arrow] (c1) -- (h1);
\draw[arrow] (c2) -- (h2);
\draw[arrow] (c3) -- (h3);
\draw[arrow] (c4) -- (h4);

\draw[arrow] (h1) -- (h12);
\draw[arrow] (h2) -- (h12);
\draw[arrow] (h3) -- (h34);
\draw[arrow] (h4) -- (h34);

\draw[arrow] (h12) -- (root);
\draw[arrow] (h34) -- (root);

\draw[arrow] (root) -- (acc);
\draw[arrow] (acc) -- (sig);

\node[draw, dashed, rounded corners, fit=(c1) (c4), inner sep=0.2cm, label={[font=\footnotesize]above:System Components}] {};
\node[draw, dashed, rounded corners, fit=(h1) (h4) (h12) (h34) (root), inner sep=0.2cm, label={[font=\footnotesize]right:Merkle Tree}] {};

\node[font=\footnotesize, text width=2.5cm] at (-1.5,-2) {\small 1. Hash components};
\node[font=\footnotesize, text width=2.5cm] at (-1.5,-4) {\small 2. Build Merkle tree};
\node[font=\footnotesize, text width=2.5cm] at (-1.5,-6) {\small 3. Commit to accumulator};
\node[font=\footnotesize, text width=2.5cm] at (-1.5,-7.5) {\small 4. Create signature};
\end{tikzpicture}
}
\caption{A Generic Composable Attestation Construction}
\label{fig:integrated-construction}
\end{figure}

\subsection{Merkle Tree-Based Construction}
Merkle tree \cite{merkle1987} satisfies composable attestation system feature requirements due to their efficient aggregation and verification capabilities, inherently supporting order independence (with proper construction) and scalable security . The process is outlined in the Merkle Tree Algorithm \ref{alg:merkle_tree}. Their use in dynamic scenarios, like certificate updates, has been well-explored \cite{naor1999merkle}.

\begin{algorithm}
\caption{Merkle Tree Proof Generation}
\label{alg:merkle_tree}
\begin{algorithmic}[1]
\REQUIRE Set of components $S = \{C_1, \ldots, C_n\}$, Hash function $H$
\ENSURE Merkle tree root $\pi_{root}$
\STATE Initialize list of leaf nodes $L = []$
\FOR{each component $C_i \in S$}
    \STATE $h_i \leftarrow H(C_i)$
    \STATE Append $h_i$ to $L$
\ENDFOR
\IF{order independence is required via sorting}
    \STATE Sort $L$ based on hash values
\ENDIF
\STATE Let $N \leftarrow L$ (current level of nodes)
\WHILE{$|N| > 1$}
    \STATE Initialize list $P = []$ (parent nodes)
    \FOR{$j \leftarrow 1$ to $|N|$ step $2$}
        \STATE $node_L \leftarrow N[j]$
        \IF{$j+1 \leq |N|$}
            \STATE $node_R \leftarrow N[j+1]$
        \ELSE
            \STATE $node_R \leftarrow node_L$ \COMMENT{Handle odd number of nodes}
        \ENDIF
        \STATE $p_j \leftarrow H(node_L \| node_R)$ \COMMENT{Concatenate and hash}
        \STATE Append $p_j$ to $P$
    \ENDFOR
    \STATE $N \leftarrow P$
\ENDWHILE
\STATE $\pi_{root} \leftarrow N[0]$
\RETURN $\pi_{root}$
\end{algorithmic}
\end{algorithm}

Verification involves recomputing the root using the provided components and their ordering (if sorted) and comparing it with the given attestation proof $\pi_{root}$. To achieve order independence, sorted Merkle trees, where leaf nodes are arranged lexicographically by their hash values, can be employed. Alternatively, commutative hashing functions, where $H(a\|b) = H(b\|a)$, could be utilized, though these require specialized cryptographic design.

\subsection{Cryptographic Accumulator Construction}
Cryptographic accumulators \cite{benaloh1990} offer an alternative for composable attestation, providing constant-size proofs irrespective of the number of attested components. Accumulators inherently satisfy order independence due to their commutative nature. Dynamic accumulators further allow for efficient updates and witness generation \cite{camenisch2002dynamic}. The general operations are described in Algorithm \ref{alg:accumulator}.

\begin{algorithm}
\caption{Accumulator-Based Attestation}
\label{alg:accumulator}
\begin{algorithmic}[1]
\REQUIRE Set of components $S = \{C_1, \ldots, C_n\}$, Accumulator functions: Initialize $AccInit()$, Add $AccAdd(acc, C)$, Witness $AccWit(acc, C)$, Verify $AccVerify(acc, C, wit)$
\ENSURE Accumulator value $\pi_{acc}$ representing $S$

\STATE $\textit{acc\_val} \leftarrow AccInit()$
\FOR{each component $C_i \in S$}
    \STATE $\textit{acc\_val} \leftarrow AccAdd(\textit{acc\_val}, C_i)$
\ENDFOR
\STATE $\pi_{acc} \leftarrow \textit{acc\_val}$
\RETURN $\pi_{acc}$

\vspace{0.2cm}
\STATE \textbf{Function} GenerateWitness($\pi_{acc}$, $C_k$):
\STATE \quad \textbf{return} $AccWit(\pi_{acc}, C_k)$

\vspace{0.2cm}
\STATE \textbf{Function} VerifyInclusion($\pi_{acc}$, $C_k$, $witness_k$):
\STATE \quad \textbf{return} $AccVerify(\pi_{acc}, C_k, witness_k)$
\end{algorithmic}
\end{algorithm}

In this construction, the accumulator value $\pi_{acc}$ serves as the attestation proof. Components can be added to the accumulator, and witness values facilitate efficient verification of individual component inclusion. RSA accumulators and bilinear map-based accumulators are examples that fulfill composability, order independence, and inclusion properties while yielding constant-size proofs.

\subsection{Multi-Signature Based Construction}
Multi-signature schemes permit the aggregation of signatures from multiple entities (or on multiple components) into a single, compact signature, delivering both efficiency and robust cryptographic assurances \cite{boldyreva2003threshold}. Algorithm \ref{alg:multi_signature} outlines this approach. BLS signatures  are a notable example offering efficient aggregation \cite{boneh2001short}.

\begin{algorithm}
\caption{Multi-Signature Based Attestation}
\label{alg:multi_signature}
\begin{algorithmic}[1]
\REQUIRE Set of components $S = \{C_1, \ldots, C_n\}$, Signing key(s) $SK$, Verification key(s) $VK$, Signing function $Sign(SK, M)$, Aggregation function $AggSig(\sigma_1, \ldots, \sigma_n)$, Verification function $VerifySig(VK, M_{agg}, \pi_{sig})$
\ENSURE Aggregated signature $\pi_{sig}$

\STATE Initialize list of individual signatures $\Sigma = []$
\FOR{each component $C_i \in S$}
    \STATE $\sigma_i \leftarrow Sign(SK_i, H(C_i))$ \COMMENT{$SK_i$ could be unique per component or a master key}
    \STATE Append $\sigma_i$ to $\Sigma$
\ENDFOR
\STATE $\pi_{sig} \leftarrow AggSig(\sigma_1, \ldots, \sigma_n)$
\RETURN $\pi_{sig}$

\vspace{0.2cm}
\STATE \textbf{Function} VerifyAttestation($VK_{agg}$, $S_{hashes}$, $\pi_{sig}$):
\COMMENT{$S_{hashes}$ is the set of component hashes}
\STATE \quad \textbf{return} $VerifySig(VK_{agg}, S_{hashes}, \pi_{sig})$
\end{algorithmic}
\end{algorithm}

Each component (or its hash) is "signed" using a cryptographic commitment or a digital signature. These individual signatures are then aggregated into a single proof $\pi_{sig}$. Verification involves checking all components against this aggregated signature. Schemes such as BLS (Boneh-Lynn-Shacham) signatures provide efficient aggregation and possess the requisite mathematical properties for composability and order independence in the aggregation process.

\subsection{Integrated Merkle-Accumulator-Signature Construction}
A particularly robust composable attestation scheme integrates Merkle trees, accumulators, and signatures into a unified construction, as depicted in Figure \ref{fig:integrated-construction}. This involves organizing components within a Merkle tree \cite{merkle1987}, committing the Merkle root to an accumulator \cite{benaloh1990} for a constant-size representation, and subsequently signing the accumulator value using a multi-signature scheme \cite{boneh2001short} for enhanced security and non-repudiation. This integrated methodology offers hierarchical verification (individual, batch, or system-wide), potential for selective disclosure of attested parts, a compact final attestation proof irrespective of system size, and security derived from multiple cryptographic hardness assumptions, potentially leveraging advanced cryptographic primitives \cite{gentry2009}. The general procedure is outlined in Algorithm \ref{alg:integrated_construction}.

\begin{algorithm}
\caption{Integrated Merkle-Accumulator-Signature Attestation}
\label{alg:integrated_construction}
\begin{algorithmic}[1]
\REQUIRE Set of components $S = \{C_1, \ldots, C_n\}$, Hash function $H$, Accumulator functions (as in Alg. \ref{alg:accumulator}), Multi-Signature functions (as in Alg. \ref{alg:multi_signature})
\ENSURE Final attestation proof $\pi_{final}$

\STATE \COMMENT{1. Merkle Tree Construction}
\STATE $\pi_{root} \leftarrow \text{MerkleTreeProofGeneration}(S, H)$ (Using Algorithm \ref{alg:merkle_tree})

\STATE \COMMENT{2. Accumulator Commitment}
\STATE $\textit{acc\_val} \leftarrow AccInit()$
\STATE $\textit{acc\_val} \leftarrow AccAdd(\textit{acc\_val}, \pi_{root})$
\STATE $\pi_{acc} \leftarrow \textit{acc\_val}$

\STATE \COMMENT{3. Signature Authentication}
\STATE $\sigma_{acc} \leftarrow Sign(SK_{system}, H(\pi_{acc}))$
\STATE $\pi_{final} \leftarrow (\pi_{root}, \text{witnesses for } \pi_{root} \text{ in } \pi_{acc}, \sigma_{acc})$ \COMMENT{Proof components may vary}

\RETURN $\pi_{final}$
\end{algorithmic}
\end{algorithm}

\section{Analysis of Properties Across Constructions}

The presented cryptographic constructions fulfill the core properties of composable attestation, albeit through differing mechanisms.

\textbf{Composability} is achieved in Merkle trees through techniques for combining or merging tree roots; in accumulators via their update functions which incorporate new elements into the existing accumulated value \cite{camenisch2002dynamic}; and in multi-signatures through the aggregation of individual signatures into a single collective signature \cite{boldyreva2003threshold}.

\textbf{Transitivity} is maintained due to the underlying cryptographic primitives. For hash-based constructions like Merkle trees, the properties of cryptographic hash functions (e.g., collision resistance, preimage resistance based on models like \cite{bellare1993random}) ensure that if a combined hash is valid, the individual components contributing to it are implicitly part of that valid state. In accumulator constructions, the membership-testing property, which allows verification that an element is part of the accumulated set \cite{benaloh1990}, inherently supports transitivity. For signature-based schemes, the correctness of the signature verification algorithm ensures that if an aggregated signature is valid over a set of messages, then the system it represents (including subsets) is considered valid under that signature.

The \textbf{efficiency of inclusion} varies significantly. Merkle trees typically offer $O(\log n)$ complexity for generating inclusion proofs \cite{naor1999merkle}. Accumulators can provide $O(1)$ size proofs for inclusion, although updating the accumulator itself might take $O(1)$ or potentially $O(n)$ time depending on the specific accumulator scheme and whether witnesses for other elements need to be updated \cite{camenisch2002dynamic}. Multi-signatures achieve $O(1)$ aggregation for some schemes (like BLS \cite{boneh2001short}), but verification might require processing all $n$ components or their hashes if individual verification is not supported by the aggregate.

Regarding \textbf{security}, each construction derives its guarantees from different cryptographic assumptions. Merkle trees rely on the collision resistance of the underlying hash function \cite{merkle1987, bellare1993random}. Accumulators' security is often based on assumptions such as the strong RSA assumption or the hardness of problems in bilinear groups \cite{benaloh1990}. Multi-signature schemes derive their security from assumptions like the discrete logarithm problem or elliptic curve discrete logarithm problem \cite{boneh2001short, boldyreva2003threshold}. The overall system security benefits from diverse cryptographic foundations \cite{gentry2009}.

\section{Applications in Distributed Systems and AI Computation}

Composable attestation offers significant advantages for ensuring integrity and trustworthiness in complex distributed systems \cite{eldib2020comprehensive}, with particular relevance to AI computation environments.

\subsection{Attestation in AI Systems}
\begin{figure}[H]
    \centering
    \resizebox{\columnwidth}{!}{%
    \begin{tikzpicture}[node distance=1.2cm, auto]
        \node[rectangle, draw, rounded corners, fill=gray!20] (modelupdate) {AI Model Update Initiation};
        \node[rectangle, draw, below of=modelupdate, fill=blue!10] (integrity) {Integrity Check via TPM/TEE};
        \node[rectangle, draw, below of=integrity, fill=blue!10] (measurements) {Collect Integrity Measurements};
        \node[rectangle, draw, below of=measurements, fill=orange!10] (attestationagent) {Attestation Agent};
        \node[rectangle, draw, below of=attestationagent, fill=orange!10] (cryptoproofs) {Generate Cryptographic Proofs};
        \node[rectangle, draw, below of=cryptoproofs, fill=green!10] (aggregation) {Composable Attestation (Merkle Trees)};
        \node[rectangle, draw, below of=aggregation, fill=green!10] (verification) {Aggregate and Verify Proofs};
        \node[rectangle, draw, below of=verification, fill=purple!10] (trustreport) {Unified Trust Report};

        \draw[->, thick] (modelupdate) -- (integrity);
        \draw[->, thick] (integrity) -- (measurements);
        \draw[->, thick] (measurements) -- (attestationagent);
        \draw[->, thick] (attestationagent) -- (cryptoproofs);
        \draw[->, thick] (cryptoproofs) -- (aggregation);
        \draw[->, thick] (aggregation) -- (verification);
        \draw[->, thick] (verification) -- (trustreport);

        \node[right=1cm of integrity] {Trusted component validation};
        \node[right=1cm of cryptoproofs] {Proofs based on cryptographic primitives};
        \node[right=1cm of aggregation] {Efficient aggregation using Merkle Trees};
        \node[right=1cm of verification] {Dynamic verification across system};
    \end{tikzpicture}%
    }
    \caption{AI Model Update Composable Attestation Process}
    \label{fig:ai-tee-attestation}
\end{figure}
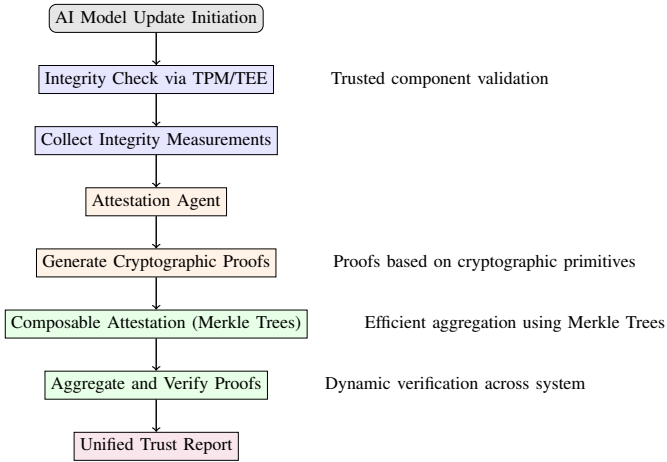

AI systems often comprise numerous interconnected components, including datasets, preprocessing modules, model architectures, learned parameters, and deployment environments. Composable attestation can provide fine-grained, verifiable integrity for each element and the overall AI pipeline.  Figure \ref{fig:ai-tee-attestation} illustrates an AI system  within a Trusted Execution Environment (TEE) \cite{intel_sgx_attestation, arm_cca_attestation}, where composable attestation can facilitate dynamic AI model updates and attestation of data and programs \cite{zhang2021attestation}. The integrity of these systems can be foundational to achieving higher-level goals like fairness in AI \cite{hardt2016equality}.

For instance, in Large Language Model (LLM) attestation, proofs can cover distinct aspects such as the execution environment ($A_{env}$), hardware security status ($A_{hw}$) derived from mechanisms like TPMs \cite{pfi_attestation_overview} or CPU-based attestation \cite{anati2013innovative}, model weight and parameter integrity ($A_{model}$), and the authenticity of software libraries and dependencies ($A_{lib}$). The overall attestation for an LLM system can then be constructed as $A_{LLM} = A(A_{env} \cup A_{hw} \cup A_{model} \cup A_{lib})$, where each component attestation is aggregated using a composable proof structure. This modularity is crucial for verifying the integrity of complex AI models where individual components may be sourced or updated independently.

In federated learning scenarios, composable attestation facilitates the verification of client contributions without compromising private data, allows for the aggregation of integrity proofs for model updates from diverse participants \cite{bonawitz2017practical}, and supports dynamic participation by efficiently including attestations from new clients. The security of such systems is an active area of research \cite{chen2019shielding}. Furthermore, it enables the verification of the entire AI model supply chain, from training data provenance and integrity, through training process attestation and model evolution tracking, to deployment environment verification. This comprehensive approach ensures end-to-end trustworthiness.

\subsection{Runtime Attestation and Dynamic System Verification}
Composable attestation is well-suited for continuous verification in dynamic environments. Incremental updates allow system components to be modified with minimal attestation overhead, as only the changed components and their linkage to the overall system state need re-attestation. This is depicted in Figure \ref{fig:incremental-attestation}, where an updated attestation $\pi'$ is formed by composing the initial attestation $\pi_0$ with the attestation of a new component $\pi_4$, without re-attesting existing components. This efficiency draws from principles in dynamic structures \cite{naor1999merkle, camenisch2002dynamic} and extends to attesting system state transitions and maintaining attestation continuity for computations that span multiple execution environments.

\begin{figure*}[h!]
\centering
\scalebox{0.8}{%
\begin{tikzpicture}[
    node distance=0.8cm,
    box/.style={draw, rectangle, minimum width=3cm, minimum height=1.2cm, align=center, font=\small},
    comp/.style={draw, rectangle, rounded corners, minimum width=3cm, minimum height=1.2cm, align=center, fill=blue!10, font=\small},
    newcomp/.style={draw, rectangle, rounded corners, minimum width=3cm, minimum height=1.2cm, align=center, fill=green!10, font=\small},
    proof/.style={draw, ellipse, minimum width=3.2cm, minimum height=1.4cm, fill=yellow!10, font=\small},
    arrow/.style={->, >=stealth, thick}
]

\node[proof] (initial) at (0,0) {Initial Attestation $\pi_0$};

\node[comp] (c1) at (-4,0) {Component $C_1$};
\node[comp] (c2) at (-4,-2) {Component $C_2$};
\node[comp] (c3) at (-4,-4) {Component $C_3$};

\draw[arrow] (c1) -- (initial);
\draw[arrow] (c2) -- (initial);
\draw[arrow] (c3) -- (initial);

\node[newcomp] (c4) at (0,-4) {New Component $C_4$};
\node[proof] (c4proof) at (4,-4) {Attestation $\pi_4$};
\draw[arrow] (c4) -- (c4proof);

\node[proof] (updated) at (8,0) {Updated $\pi'$};

\draw[arrow] (initial) -- (updated);
\draw[arrow] (c4proof) -- node[above, sloped, font=\scriptsize] {\normalsize Compose} (updated);

\node[box, fill=orange!10] (v1) at (4,2.5) {Verify};
\node[box, fill=green!10] (v2) at (11,2.5) {Verify};

\draw[arrow] (initial) -- (v1);
\draw[arrow] (v1) -- node[above, font=\scriptsize] {\normalsize Valid} (v2);
\draw[arrow] (updated) -- (v2);

\node[align=center, font=\small] at (0,2) {Initial System};
\node[align=center, font=\small] at (8,2) {Updated System};

\node[align=center, text width=12cm, font=\scriptsize] at (4,-6) { \normalsize
The updated attestation $\pi'$ is computed by composing the initial attestation $\pi_0$ with the new component's attestation $\pi_4$ without re-attesting components $C_1$, $C_2$, and $C_3$.\\
$\pi' = \pi_0 \oplus \pi_4$
};
\end{tikzpicture}}
\caption{Incremental Attestation with Composable Construction (Enhanced for Visibility).}
\label{fig:incremental-attestation}
\end{figure*}

The choice of attestation construction can be tailored to specific application requirements. High-performance computing systems might favor Merkle tree-based constructions for their efficient verification processes. Applications demanding strong privacy, such as those handling sensitive data in AI model training or inference, could benefit from constructions that can integrate with zero-knowledge proof principles or secure multi-party computation \cite{lindell2017secure}, even if not detailed as a standalone construction herein. Highly dynamic systems, where components are frequently added or updated, may find accumulator-based constructions advantageous due to their potential for efficient updates. Decentralized systems often align well with multi-signature approaches, which mirror distributed trust models by allowing multiple parties to contribute to the overall system attestation. The goal often is to achieve a form of property-based attestation \cite{sadeghi2015property} relevant to the application's needs.

\section{Conclusion and Future Work}
This paper has introduced a generalized framework for composable attestation, formalizing the mathematical properties essential for scalable, modular, and cryptographically secure integrity verification in distributed systems. By exploring multiple cryptographic constructions, including Merkle trees, accumulators, and multi-signature schemes, we have demonstrated that composable attestation can be realized through diverse mechanisms, each offering distinct trade-offs in performance, security, and functionality. The inherent property of inclusion further enhances composable attestation by permitting incremental proof updates, thereby diminishing computational overhead in dynamic settings. The framework's applicability to AI systems underscores its versatility and significance in establishing trust within intricate, distributed computational landscapes, building on the fundamental need to bootstrap trust in computing systems \cite{parno2011bootstrapping}.

Future research endeavors could focus on several promising directions. The development of advanced hybrid attestation constructions that synergistically combine multiple cryptographic approaches may yield enhanced security and efficiency. Optimizations tailored for real-time AI attestation in highly dynamic environments warrant further investigation. Exploring privacy-preserving attestation techniques suitable for sensitive AI applications remains a critical area, potentially drawing from fields like secure multi-party computation \cite{lindell2017secure}. Additionally, the formal verification of specific attestation implementations against the defined properties, similar to efforts in verifying systems like DICE \cite{vasudevan2021dice} or general concurrent systems \cite{hunt2020verifying}, would bolster confidence in their correctness and security, ideally within strong formal frameworks like UC \cite{cai2021universal}. Finally, standardization efforts \cite{pfi_attestation_overview, arm_cca_attestation, intel_sgx_attestation} are crucial for fostering interoperable attestation mechanisms across disparate systems and platforms, which will be vital as AI and distributed systems continue to proliferate and interconnect. As these systems grow in complexity and distribution, composable attestation provides an indispensable foundation for establishing and maintaining trust through robust, cryptographically verifiable evidence.

\end{document}